\documentclass[aps,prc,twocolumn,showpacs,nofootinbib]{revtex4-1}
\usepackage[utf8]{inputenc}
\usepackage{graphicx}   
\usepackage{latexsym}   
\usepackage{enumerate}
\usepackage{rotating,booktabs,multirow}
\usepackage{amsmath}
\usepackage{amsfonts}
\usepackage{amssymb}
\usepackage{multirow}
\usepackage{bm}
\usepackage{lipsum}
\usepackage{xcolor}

\usepackage{soul}

\begin{document}
\title{Comparison of heavy ion transport simulations: Ag+Ag collisions at $E_\mathrm{lab}=1.58$~$A$GeV}
\author{Tom Reichert$^{1,6}$, Alexander Elz$^{1}$, Taesoo Song$^{2}$, Gabriele Coci$^{2}$, Michael~Winn$^{4}$, Elena~Bratkovskaya$^{2,1,6}$, J\"org~Aichelin$^{4,5}$, Jan~Steinheimer$^{5}$, Marcus~Bleicher$^{1,2,3,6}$}

\affiliation{$^1$ Institut f\"ur Theoretische Physik, Goethe Universit\"at Frankfurt, Max-von-Laue-Strasse 1, D-60438 Frankfurt am Main, Germany}
\affiliation{$^2$ GSI Helmholtzzentrum f\"ur Schwerionenforschung GmbH, Planckstr. 1, 64291 Darmstadt, Germany}
\affiliation{$^3$ John von Neumann-Institut f\"ur Computing, Forschungzentrum J\"ulich,
52425 J\"ulich, Germany}
\affiliation{$^4$ SUBATECH UMR 6457 (IMT Atlantique, Universit\'e de Nantes, IN2P3/CNRS), 4 Rue Alfred Kastler, F-44307 Nantes, France}
\affiliation{$^5$ FIAS, Ruth-Moufang-Str.1, D-60438 Frankfurt am Main, Germany}
\affiliation{$^6$ Helmholtz Research Academy Hesse for FAIR (HFHF), GSI Helmholtz Center for Heavy Ion Physics, Campus Frankfurt, Max-von-Laue-Str. 12, 60438 Frankfurt, Germany}

\begin{abstract}
We compare the microscopic transport models UrQMD, PHSD, PHQMD, and SMASH to make predictions for the upcoming Ag+Ag data at $E_\mathrm{lab}=1.58$~$A$GeV ($\sqrt{s_\mathrm{NN}}=2.55$~GeV) by the HADES collaboration. We study multiplicities, spectra and effective source temperatures of protons, $\pi^{\pm,0}$, $K^\pm$, the $\eta$, $\Lambda+\Sigma^0$ and the $\Xi^-$ within these models. Despite 
variations in the detailed implementation of the dynamics in the different models, the employed transport approaches all show consistent multiplicities of the bulk of investigated hadrons. The main differences are in the $\Xi^-$ production, which is treated differently between UrQMD/SMASH on one side employing high mass resonance states with explicit decays to $\mathrm{Resonance}\rightarrow \Xi+K+K$ in contrast to PHSD/PHQMD which 
account only non-resonant $\Xi$ production channels.
A comparison of the spectra, summarized by effective source temperatures, shows that all models provide similar source temperatures around $T_\mathrm{source}=80-95$~MeV, and show substantial radial flow on the order of $\langle v_T\rangle=0.22c-0.3c$ even for such a small system.
\end{abstract}

\maketitle

\section{Introduction}
Over the last two decades, collider facilities such as the Relativistic Heavy-Ion Collider (RHIC) located at Brookhaven National Laboratory (BNL) or the Large Hadron Collider (LHC) at European Organization for Nuclear Research (CERN) have studied the strongly interacting matter created in heavy-ion collisions at relativistic energies. Here, highest temperatures and an almost net baryon densities free ($\mu_\mathrm{B}\approx 0$) system was created with a deconfined initial state, called the Quark-Gluon Plasma (QGP).
On the other hand, research at GSI Helmholtzzentrum für Schwerionenforschung (GSI) in Darmstadt, Germany and the upcoming Facility for Antiproton and Ion Research (FAIR) in Darmstadt, Germany and the Nuclotron-based Ion Collider fAcility (NICA) in Dubna, Russia aims to explore the QCD phase diagram in the region of very high baryon densities (a similar program is also active at BNL with the beam energy scan program). 
The primary focus of these programs is to gain further insight into overarching questions such as: the equation of state of nuclear matter, in-medium modifications of hadrons and their resonances and at higher energies locating the critical end point of the phase transition between hadronic and quark-gluon matter and the onset of chiral symmetry restoration.
At very low beam energies, this high-$\mu_\mathrm{B}$ region of the QCD phase diagram is currently explored by the High Acceptance Di-Electron Spectrometer (HADES) experiment \cite{Agakishiev:2009am} at GSI which has firmly established its abilities to precisely measure di-leptons and constrain vector-meson spectral functions \cite{Agakichiev:2006tg,Agakishiev:2011vf}, to determine event-by-event correlations and fluctuations \cite{Adamczewski-Musch:2020slf}, to investigate final state n-particle correlations \cite{Adamczewski-Musch:2020edy} and to measure rare probes and sub-threshold strangeness production \cite{Agakishiev:2009rr,Agakishiev:2010rs,Adamczewski-Musch:2017rtf}. Recently, HADES has reported first preliminary data on hadron production and di-lepton measurements in silver-silver (Ag+Ag) collisions at $E_\mathrm{lab}=1.58$~$A$GeV ($\sqrt{s_\mathrm{NN}}=2.55$~GeV) \cite{Szewczyk:2020pyk,HADES:2020nme}. This will allow to pin-down the properties of dense hadronic matter further and complements the data on Au+Au reactions taken in the past.

From a theoretical perspective, there are several different approaches to describe experimental results of heavy ion collisions, which all rely on different assumptions: I) Thermal models \cite{Schnedermann:1993ws,Cleymans:1999st,Andronic:2017pug} that assume global thermodynamic equilibrium and provide information on the integrated abundances of hadrons, II) hydrodynamic approaches \cite{Sollfrank:1996hd,Huovinen:2001cy,Romatschke:2007mq,Niemi:2015qia} assuming local thermal equilibrium, that emphasize the collective, fluid-like behavior of the matter and allow to explore abundances and spectra and III) transport models \cite{Hartnack:1997ez,Bass:1998ca,Bleicher:1999xi,Cassing:2009vt,Bratkovskaya:2011wp,Buss:2011mx,Weil:2016zrk} that do not rely on equilibrium assumptions, but follow from the kinetic theory and allow to study yields, spectra and dynamical fluctuations and correlations.  

While the definition of the physics in the thermal and hydrodynamical models is rather straightforward (either given by the partition function or the Equation-of-State (EoS)), transport approaches rely on a multitude of assumptions and models to describe the properties of individual hadrons, the hadronic cross sections, the potentials (which reflect the EoS) or the onset of string formation. Also differences in the numerical implementations of similar assumptions are possible, e.g. parallel ensemble simulations vs. full ensemble simulations. Thus, it is a priori not clear, if the very similar physical assumptions implemented in the various approaches provide quantitatively similar results. For a systematic study of the general uncertainties in such models we refer the reader to \cite{Gerhard:2012fj,Ono:2019ndq}.

Motivated by upcoming HADES data of Ag+Ag collisions at 1.58~$A$GeV kinetic beam energy, we present a comparative analysis of bulk properties of $p$, $\pi^{\pm,0}$, $K^\pm$, $\eta$, $\Lambda+\Sigma^0$ and $\Xi^-$ hadrons employing the most prominent transport approaches used at SIS energies, namely UrQMD \cite{Bass:1998ca,Bleicher:1999xi}, PHSD \cite{Cassing:2009vt,Bratkovskaya:2011wp}, PHQMD \cite{Aichelin:2019tnk} and SMASH \cite{Weil:2016zrk}. We compare particle production mechanisms and the different Equations-of-State via total multiplicities, rapidity and transverse mass distributions and extract an effective temperature $T_\mathrm{eff}$ from the transverse mass spectra. 

\section{Short description of the transport models}

All employed transport models for this study have been well documented in various publications. Here, we shortly summarize the main features of these approaches. While the models discussed here generally fall into two categories, that mainly denote how the underlying relativistic transport equations are numerically solved, i.e.
\begin{itemize}
    \item event based Quantum Molecular Dynamics (QMD) type approaches and
    \item density based (Vlasov) Boltzmann-Uehling-Uhlenbeck ((V)BUU) type approaches.
\end{itemize}
In both cases the general physics assumptions and results are very similar. Main differences are related to different physical choices, e.g. the type of potential used at these low energies or how multi-strange baryons are produced.

\subsection{UrQMD}
UrQMD v3.5 is based on an explicit propagation of all hadrons in the reactions either in cascade mode or with a hard or soft nuclear potential. The collision term contains 70 baryon- and 39 meson-species and their corresponding antiparticles.
In UrQMD the colliding nuclei are modeled either as superpositions of gaussian phase space densities of each nucleon, if potentials are used or as a collection of point-like nucleons, if the cascade mode is employed. In both modes, elastic or inelastic reactions are treated by means of their geometrical cross-section $\sigma_\text{tot}=\sigma_\text{inelastic}+\sigma_\text{elastic}$, decays proceed via measured branching ratios. A collision takes place, when the transverse distance $d_\text{trans}$ between the two approaching hadrons becomes less than
\begin{equation}
d_\text{trans}\leq d_0=\sqrt{\frac{\sigma_\text{tot}}{\pi}}\, .
\label{eq:collcrit}
\end{equation}

The outgoing particles of such a binary collision then interact further creating the evolution of the system. Thus, UrQMD provides a microscopic implementation of the scattering term with strict conservation of energy, momentum and all quantum numbers. For details of the model we refer the reader to \cite{Graef:2014mra,Bass:1998ca,Bleicher:1999xi}

\subsection{PHSD}
The Parton-Hadron-String Dynamics (PHSD) \cite{Cassing:2008sv,Cassing:2008nn,Cassing:2009vt,Bratkovskaya:2011wp,Linnyk:2015rco,Moreau:2019vhw} is a microscopic covariant  transport approach for the dynamical  description of strongly interacting  hadronic and partonic matter created in heavy-ion collisions. 
It is based on a solution of the Cassing-Juchem generalised off-shell transport equations for test particles \cite{Cassing:1999wx,Cassing:1999mh}, which are derived from the  Kadanoff-Baym equations  in first-order gradient expansion \cite{Juchem:2004cs,Cassing:2008nn}.
This quantum field theoretical basis distinguishes the PHSD from the semi-classical BUU based model, since the PHSD  propagates Green's functions (in phase-space representation) which contain information not only on the occupation probability (in terms of the phase-space distribution functions), but also on the properties of hadronic and partonic degrees-of-freedom via their spectral functions. The PHSD approach consistently describes the full evolution of a relativistic heavy-ion collision from the initial hard scatterings and string formation through the dynamical deconfinement phase transition 
to the strongly-interacting quark-gluon plasma (sQGP) as well as hadronization and the subsequent interactions in the expanding hadronic phase as in the Hadron-String-Dynamics (HSD) transport approach \cite{Ehehalt:1996uq,Cassing:1999es}. For details of the model we refer the reader to \cite{Bratkovskaya:2011wp,Linnyk:2015rco,Moreau:2019vhw}.

We note that for this study we employed the PHSD v.~4.5 which includes the modification of the properties of strange degrees-of-freedom  in the hot and dense medium  (cf. Ref. \cite{Song:2020clw}):
the antikaon $(\bar K = K^-, \bar K^0)$ properties are described via many-body G-matrix calculations, while the in-medium modification of kaons $(K = K^+, K^0)$ are accounted via
the kaon-nuclear potential, which is assumed to be proportional to the local baryon density.

\subsection{PHQMD}
The PHQMD (Parton-Hadron-Quantum-Molecular-Dynamics) approach \cite{Aichelin:2019tnk}, a branch of the PHSD, is a novel $n$-body dynamical transport approach  which is designed to provide a microscopic description of nuclear cluster and hypernucleus formation as well as of general particle production in heavy-ion reactions at relativistic energies. In difference to the coalescence or statistical models, often 
used for the cluster formation, in PHQMD clusters are formed dynamically due to the interactions between baryons described on the basis of Quantum Molecular Dynamics (QMD)
which allows to propagate the $n$-body Wigner density and $n$-body correlations in phase-space, that are essential for the cluster formation. The PHQMD provides a good description of the hadronic single particle observables \cite{Aichelin:2019tnk} as well as a dynamical cluster production from SIS to RHIC energies \cite{Aichelin:2020vpb,Glassel:2021rod}.

\subsection{SMASH}
The Simulating Many Accelerated Strongly-interacting  Hadrons (SMASH) \cite{Weil:2016zrk} is based on the test-particle set-up for the hadronic dynamics. It includes a large body of established hadrons and their resonances as degrees of freedom. The dynamics is modeled either in cascade mode or potential interactions  can be employed (here we refer only to the cascade results published in \cite{Staudenmaier:2020xqr}). At higher energies SMASH involves string degrees of freedom using the PYTHIA model. For the recent results on Ag+Ag reactions from SMASH we refer to \cite{Staudenmaier:2020xqr}.

\section{Model predictions and comparisons - general set-up}
For the present investigation, we focus on Ag+Ag collisions at $E_\mathrm{lab}=1.58$~$A$GeV. Results at other collision energies and systems have already been published. The calculations are performed at 0-10\% centrality. The centrality is determined via an impact parameter cut derived from the hard sphere approximation, i.e. $b_\mathrm{max}=3.6$~fm. Using a geometrical cut in this approximation provides generally a good description of the experimental centrality, especially in case of the HADES centrality cuts a full modeling of the centrality selection can not be achieved without a full GEANT simulation. Previous studies have shown that this approximation is generally good to level of better than 10\% . For the rest of the paper we compare UrQMD, PHSD and PHQMD results for a hard and soft Equation-of-State (in case of PHSD, the EoS with in-medium effects can be seen as in the ``middle'' between both cases, because its compressibility is in between a hard and soft EoS) and SMASH results \cite{Staudenmaier:2020xqr} in cascade mode. 

The simulation data of UrQMD, PHSD and PHQMD is summarized in the auxiliary files in the online version of this manuscript, the data from SMASH can be found in \cite{Staudenmaier:2020xqr}. 
\begin{figure}[t!]
    \centering
    \includegraphics[width=\columnwidth]{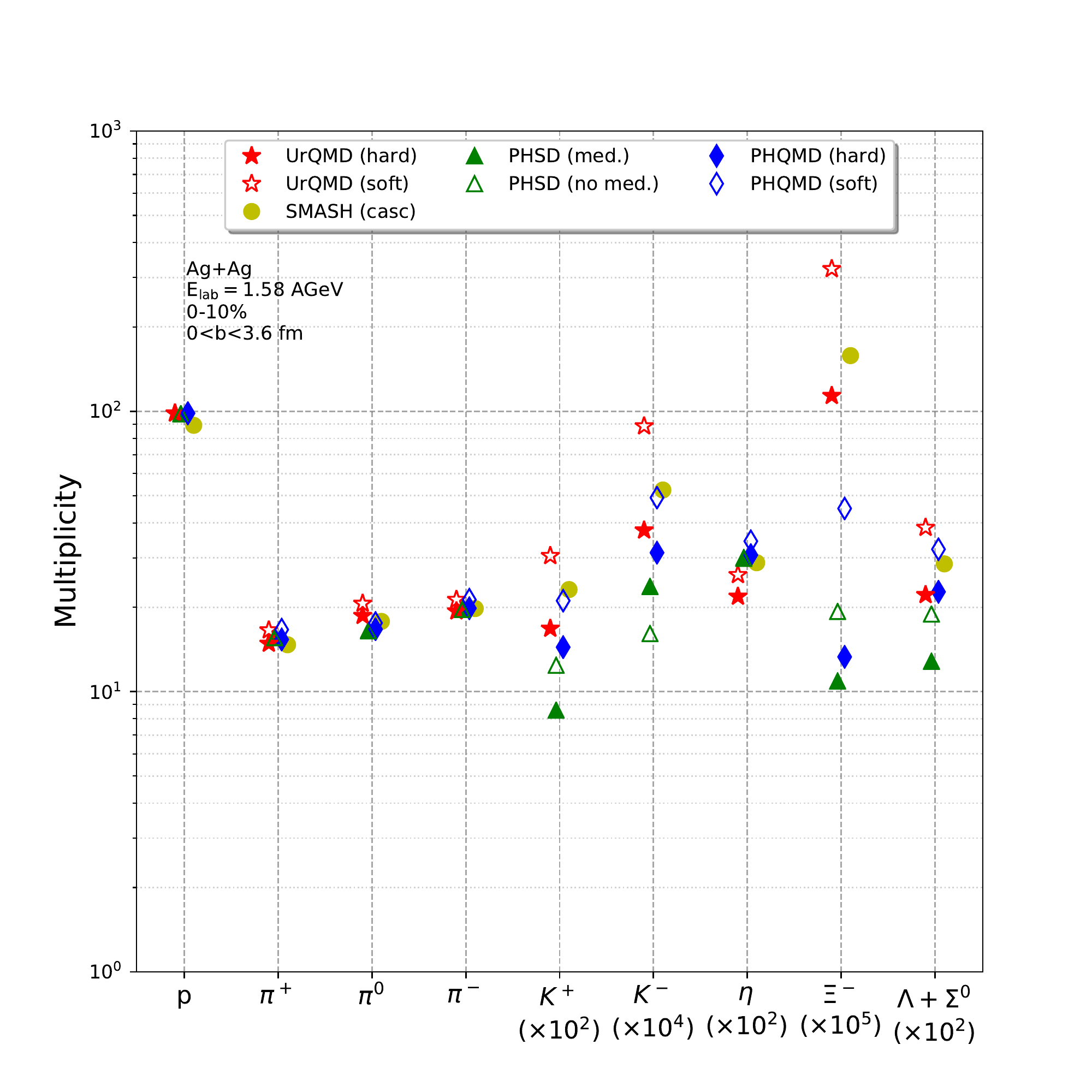}
    \caption{[Color online] Integrated hadron multiplicities for protons, pions, $K^+$ (multiplied by $10^2$), $K^-$ (multiplied by $10^4$), $\eta$ (multiplied by $10^2$), $\Lambda+\Sigma^0$ (multiplied by $10^2$) and $\Xi^-$ (multiplied by $10^5$) in 0-10\% central Ag+Ag collisions at 1.58~$A$GeV from UrQMD (red stars), SMASH \cite{Staudenmaier:2020xqr} (yellow circles), PHSD (green triangles) and PHQMD (blue diamonds). The open symbols denote the soft/no-medium interaction versions, while the full symbols denote the hard/in-medium version of the models. The cascade mode results from SMASH are also shown by full symbols.}
    \label{multiplicity}
\end{figure}

\begin{figure}[t!]
    \centering
    \includegraphics[width=\columnwidth]{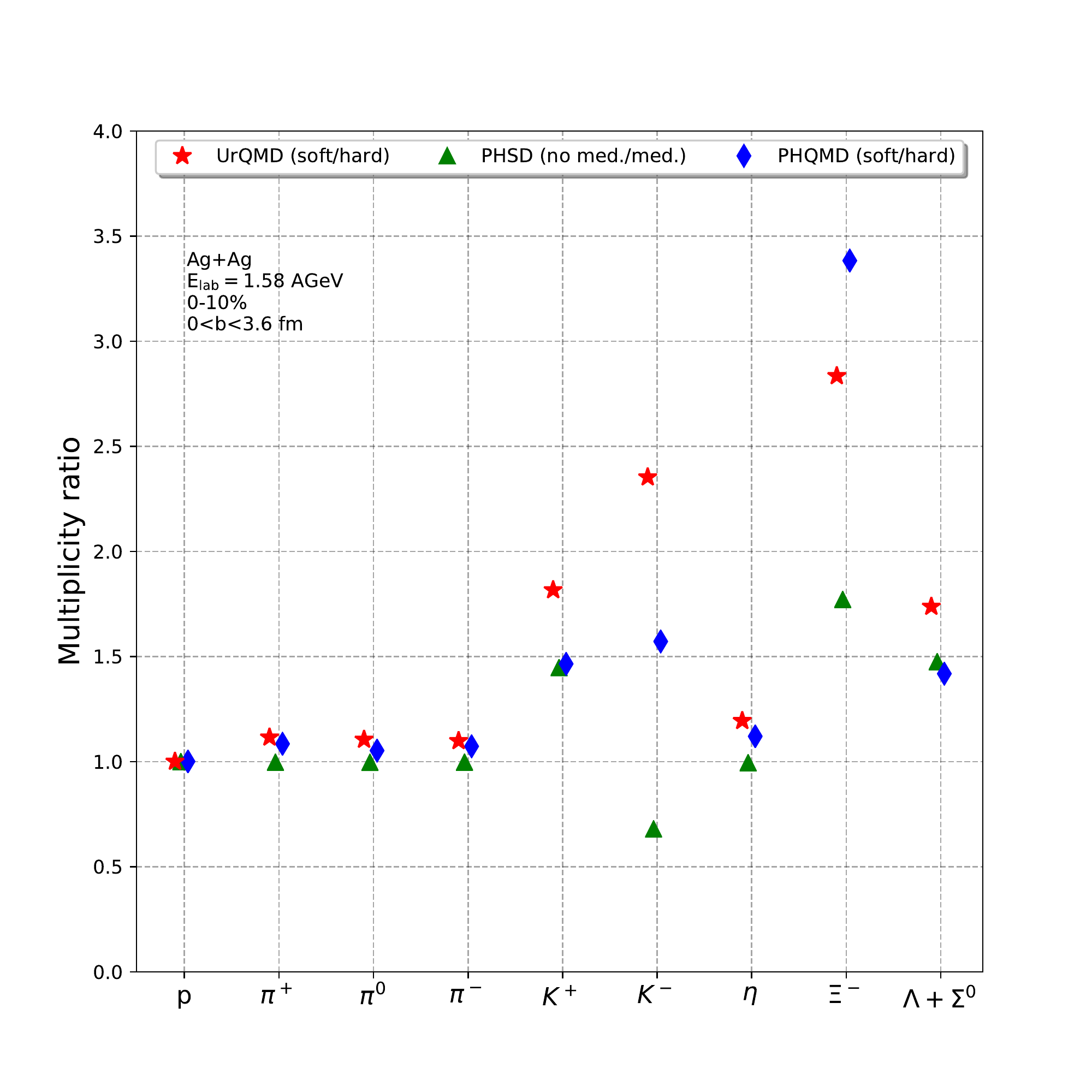}
    \caption{[Color online] Ratio (soft EoS divided by hard
EoS or in case of PHSD the ratio of calculations without medium effects to calculations with medium effects) of the integrated hadron multiplicities for protons, pions, $K^+$, $K^-$, $\eta$, $\Lambda+\Sigma^0$ and $\Xi^-$ in 0-10\% central Ag+Ag collisions at 1.58~$A$GeV from UrQMD (red stars), PHSD (green triangles) and PHQMD (blue diamonds).}
\label{fig:multiplicity_ratio}
\end{figure}

\subsection{Integrated multiplicities}
Let us start with the investigation of the $4\pi$ hadron multiplicities. In Fig. \ref{multiplicity} we show the integrated hadron multiplicities for protons, pions, $K^+$, $K^-$, $\eta$, $\Lambda+\Sigma^0$ and $\Xi^-$ in Ag+Ag collisions at 1.58~$A$GeV from UrQMD (red stars), SMASH (yellow circles), PHSD (green triangles) and PHQMD (blue diamonds) for 0-10\% centrality. The open symbols denote the soft/no-medium interaction versions, while the full symbols denote the hard/in-medium version of the models. The cascade mode results from SMASH are also shown by full symbols. One observes that all models provide consistent multiplicities of $p$, $\pi^{\pm,0}$ and the $\eta$. However, when going to single strange hadrons such as $K^\pm$, $\Lambda+\Sigma^0$ noticeable differences among the transport models and their different modes concerning the production of strangeness emerge. 

The most striking difference between UrQMD/SMASH and PHSD/PHQMD is found for the production of the double strange $\Xi^-$ hyperons. Here the yields of UrQMD/SMASH are for all centralities a factor of approx. 10-15 higher than for the PHSD/PHQMD simulations. The main reason for this difference is due to heavy resonance decays. In \cite{Steinheimer:2015sha,Steinheimer:2016vzu} the problem of the surprisingly large production rate of $\Xi^-$ hyperons in the SIS energy regime was analyzed. At these near and sub-threshold energies, the reaction dynamics is dominated by the formation of baryon resonances and their decays. $\Xi^-$ production can be rather naturally incorporated  by the following reaction: $N+N\rightarrow N+N^*$ and the eventual decay $N^*\rightarrow\Xi^-+2K$ at a later stage of the entire collision event. Unfortunately, the branching ratios of decay channels for very heavy resonances which lead to rare and heavy particle species such as $\Xi^-$ can only be indirectly fixed to experimental data, because elementary $\Xi^-$ production data does not exist at such low energies. Therefore, in Ref. \cite{Steinheimer:2015sha,Steinheimer:2016vzu} the p+Nb data was used to fix the branching ratio. One should note that these heavy resonances are exclusively formed via secondary reactions such as $N+N^*(m_1)\rightarrow N+N^*(m_2>m_1)$. By virtue of these inelastic secondary reactions, resonant states $N^*(m_1)$ undergo further excitation into higher mass states $N^*(m_2)$ which act as an energy reservoir to facilitate sub-threshold production of heavy strange baryons such as $\Xi^-$. Thus, the inclusion of heavy baryon resonances allows lifting the production threshold known from elementary ($p+p$) collisions and enables substantial multi-strange particle production already at $E_\text{lab}$~=~1.58~$A$GeV. Both PHSD and PHQMD do not include this resonance channel and do therefore produce substantially less $\Xi$ hyperons, SMASH has followed the UrQMD approach and has also included the heavy resonance decay, however with slightly modified parameters.

Nevertheless a rather model independent trend can be observed when one looks only at the difference between a soft EoS and a hard EoS. Fig. \ref{fig:multiplicity_ratio} shows the ratios of the integrated multiplicities for a soft EoS divided by the results of the hard EoS (soft EoS divided by hard
EoS or in case of PHSD the ratio of calculations without medium effects to calculations with medium effects) for the same hadrons. Here all models 
predict a sizable increase of strange particle yields for the soft EoS. 
The general trend is, the softer the Equation-of-State, the more $K$ and $\Lambda+\Sigma^0$ and $\Xi^-$ are produced. The reason is that with a softer EoS higher densities can be reached and the expansion of the system goes slow which allows for more interactions \cite{Hartnack:2005tr} . 
However, one cannot omit that there are substantial quantitative differences between the predictions of the models at high densities for (multi-)strange particles.

\subsection{Rapidity distributions}
\begin{figure}[t!]
    \centering
    \includegraphics[width=\columnwidth]{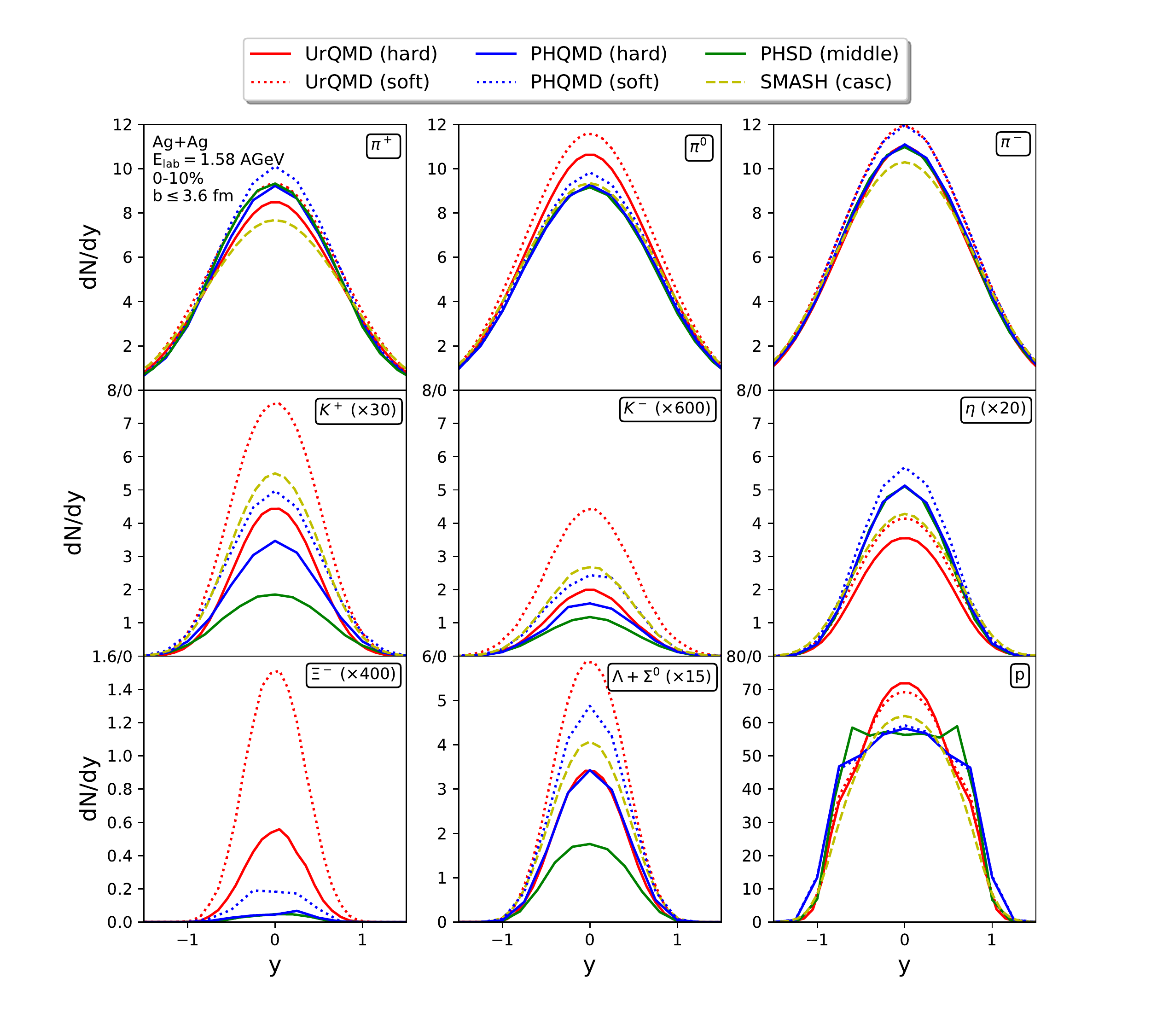}
    \caption{[Color online] Rapidity distributions of $\pi^+$ (upper left), $\pi^0$ (upper center), $\pi^-$ (upper right), $K^+$ (middle left), $K^-$ (middle center), $\eta$ (middle right), $\Xi^-$ (lower left), $\Lambda+\Sigma^0$ (lower center) and $p$ (lower right) in 0-10\% central ($b\leq3.6$~fm) Ag+Ag collisions at 1.58~$A$GeV beam energy from UrQMD (red), SMASH \cite{Staudenmaier:2020xqr} (yellow), PHSD (green) and PHQMD (blue). The dotted lines denote the soft/no-medium interaction versions, while the full lines denote the hard/in-medium version of the models and the dashed line denotes cascade mode.}
    \label{dNdy_hadrons}
\end{figure}

After the exploration of the total multiplicities we now turn in Fig. \ref{dNdy_hadrons} to the rapidity distributions. We show the rapidity densities of $\pi^+$ (upper left), $\pi^0$ (upper center), $\pi^-$ (upper right), $K^+$ (middle left), $K^-$ (middle center), $\eta$ (middle right), $\Xi^-$ (lower left), $\Lambda+\Sigma^0$ (lower center) and $p$ (lower right) in 0-10\% central ($b\leq3.6$~fm) Ag+Ag collisions at 1.58~$A$GeV beam energy from UrQMD (red), SMASH \cite{Staudenmaier:2020xqr} (yellow), PHSD (green) and PHQMD (blue). The dotted lines denote the soft/no-medium interaction versions, while the full lines denote the hard/in-medium version of the models and the dashed line denotes cascade mode.

We observe that UrQMD 3.5, PHSD, PHQMD and SMASH all provide rather similar rapidity densities for the pions as expected from the integrated yield. One also observes that the initial isospin imbalance of the silver nucleus is transported into the final state pions as expected. Also the proton distributions from all models are rather similar. We conclude from this that the energy deposition and overall particle production is similar in all investigated approaches. However, the distributions of $K^{\pm}$, $\eta$ as well as of the $\Xi^-$ and $\Lambda+\Sigma^0$ from PHSD and PHQMD exhibit distinct deviations from the distributions obtained via UrQMD, albeit PHSD and PHQMD tend to predict rather similar results compared to one another since they use a similar collision integral, except of strangeness, and differ by the propagation of the baryons by the QMD or BUU equations-of-motion. However, if all models are run with full potential interactions, this differences shrinks and the models show similar results on a level of 30\% for single strange hadrons. The larger deviation for $\Xi$'s was already discussed above and is also present in the rapidity distributions (also for the associated kaon production). 

\begin{figure*}[t!]
    \centering
    \includegraphics[width=2\columnwidth]{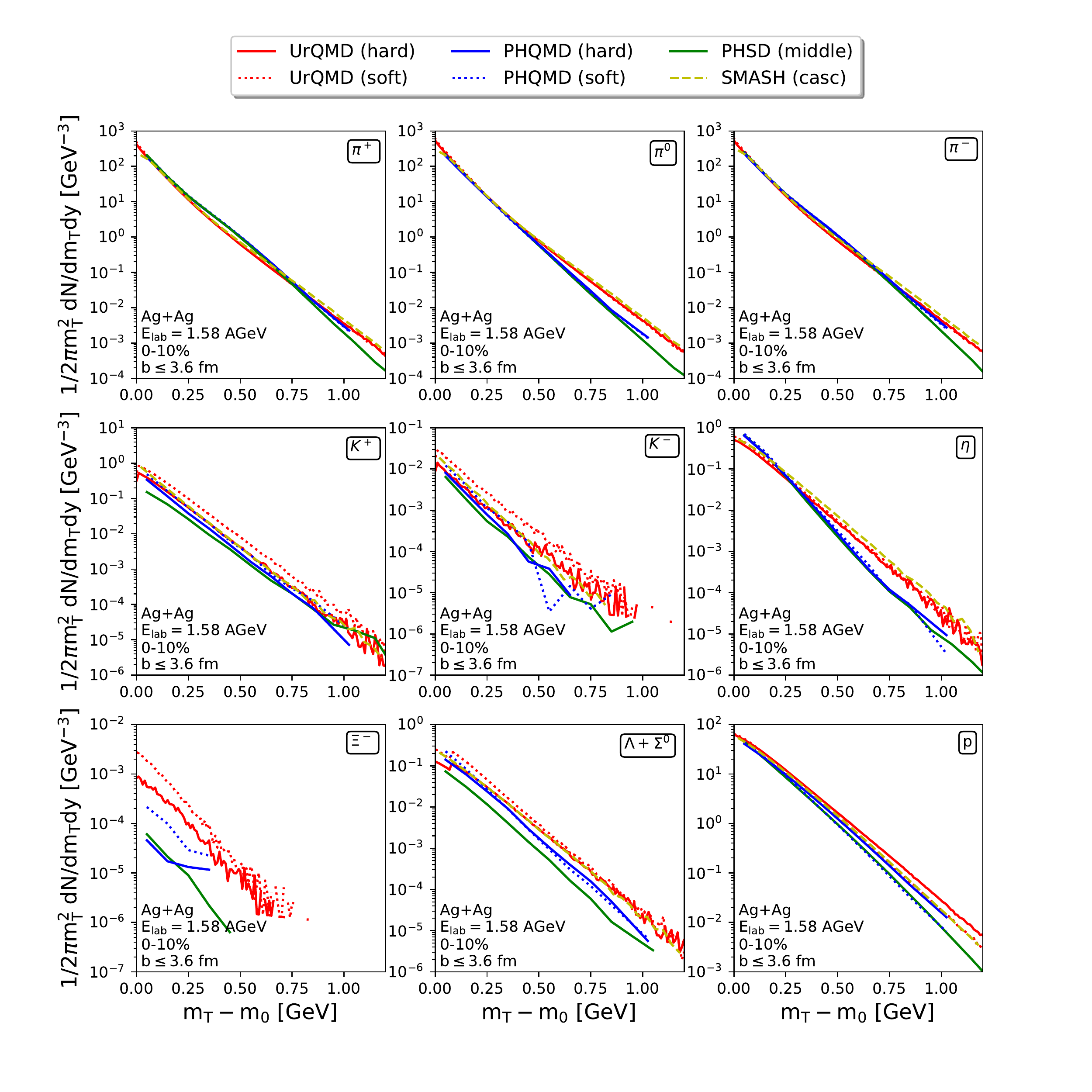}
    \caption{[Color online] Transverse mass distributions at midrapidity of $\pi^+$ (upper left), $\pi^0$ (upper center), $\pi^-$ (upper right), $K^+$ (middle left), $K^-$ (middle center), $\eta$ (middle right), $\Xi^-$ (lower left), $\Lambda+\Sigma^0$ (lower center) and $p$ (lower right) in 0-10\% central ($b\leq3.6$~fm) Ag+Ag collisions at 1.58~$A$GeV beam energy from UrQMD (red), SMASH \cite{Staudenmaier:2020xqr} (yellow), PHSD (green) and PHQMD (blue). The dotted lines denote the soft/no-medium interaction versions, while the full lines denote the hard/in-medium version of the models and the dashed line denotes cascade mode.}
    \label{fig:mT_spectra_ymid}
\end{figure*}

\subsection{Transverse mass spectra}
Next we analyze the transverse mass spectra of the investigated hadrons calculated by the different transport models. The full rapidity dependent transverse mass spectra are provided in the appendix. There we show transverse mass spectra of pions in Fig. \ref{fig:mT_spectra_pions}, the spectra of $K^{\pm}$ and $\eta$ in Fig. \ref{fig:mT_spectra_mesons} and the spectra of $p$, $\Lambda+\Sigma^0$ and $\Xi^-$ are displayed in Fig. \ref{fig:mT_spectra_baryons} in different rapidity bins (see legend) in the 10\% most central Ag+Ag collisions at $E_\mathrm{lab}=1.58A$~GeV from UrQMD v3.5, PHSD, PHQMD and SMASH. However, to focus on the differences and similarities of the models, we show in Fig. \ref{fig:mT_spectra_ymid} only the transverse mass spectra of all investigated hadrons at midrapidity. 

Let us start with the discussion of the pion distributions (top row of Fig. \ref{fig:mT_spectra_ymid}): First of all one observes that at low transverse masses ($m_{\rm T}-m_0<0.5$~GeV) all models show very similar results. This to be expected, because the low momentum pions are predominantly created from the decay of Delta-resonances. As the Delta-resonances are the dominant source of pions and all models yield similar pion yields also the Delta yields are similar and therefore the low momentum pions are similar. However, at high transverse masses the results from PHSD and PHQMD show a lower slope (less pions) as compared to all other approaches. This can be related to two sources: I) an attractive in-medium potential that slows down the high-momentum pions (also related to the slightly lower slopes of the baryons in these approaches (as will be discussed below)) and II) due to differences in the baryon resonance lists, since high mass resonance decays ($N^*\rightarrow N+\pi$) give higher pion momenta. A similar feature is also present in the case of the $\eta$ meson.

Next we turn to Kaons and anti-Kaons (middle row of Fig. \ref{fig:mT_spectra_ymid}). Here we observe that all models predict very similar slopes and also the differences between the soft and hard EoS has only minor influence (apart from the overall normalization as discussed above). In the case of the $\Lambda$ and proton transverse mass distribution (bottom row of Fig. \ref{fig:mT_spectra_ymid}), we also observe very similar slopes for all models (for Lambda's essentially independent of the EoS). For protons the expected splitting between the soft and hard EoS becomes more visible at high transverse masses. The statistics for the $\Xi$ baryons does not allow to draw any firm conclusions.

To summarize, all models show consistent transverse mass distributions for the bulk of the matter below $m_{\rm T}-m_0$ smaller than 500 MeV. At high transverse masses the effects of the EoS become visible and are mainly reflected in a hard/soft splitting of the proton distribution which also influences the high transverse mass tail of the pions. 

\subsection{Effective temperature}
Next we extract the effective temperature of the source via the rapidity dependence of the inverse slope parameters of the transverse mass spectra of the individual hadron species. For this an exponential of the form shown in Eq. \eqref{eq:exponential} is used in the transverse mass range from 0 to 800~MeV for protons, kaons, the $\eta$, $\Lambda+\Sigma^0$ and the $\Xi^-$ and in case of the pions two separate temperatures are extracted to take into account the strong Delta contribution, i.e. we extract one inverse slope in the region from $m_{\rm T}-m_0=0-300$~MeV and one for the high transverse mass tail with $m_{\rm T}-m_0=300-800$~MeV. The exponential (and the fit range) used to fit the calculations has the general form that is also used at the HADES experiment \cite{HADES:2020ver}:
\begin{equation}\label{eq:exponential}
    \frac{1}{m_{\rm T}^2}\frac{\mathrm{d}^2N}{\mathrm{d}m_\mathrm{T}\mathrm{d}y} = C(y)\exp\left(-\frac{m_\mathrm{T}}{T_\mathrm{inv}(y)}\right),
\end{equation}
where $m_\mathrm{T}$ and $m_0$ are the transverse and the rest mass, respectively, $T_\mathrm{inv}(y)$ is the rapidity-dependent inverse slope parameter and $C(y)$ is a rapidity dependent normalization factor.

\begin{figure} [t!hb]
    \includegraphics[width=\columnwidth]{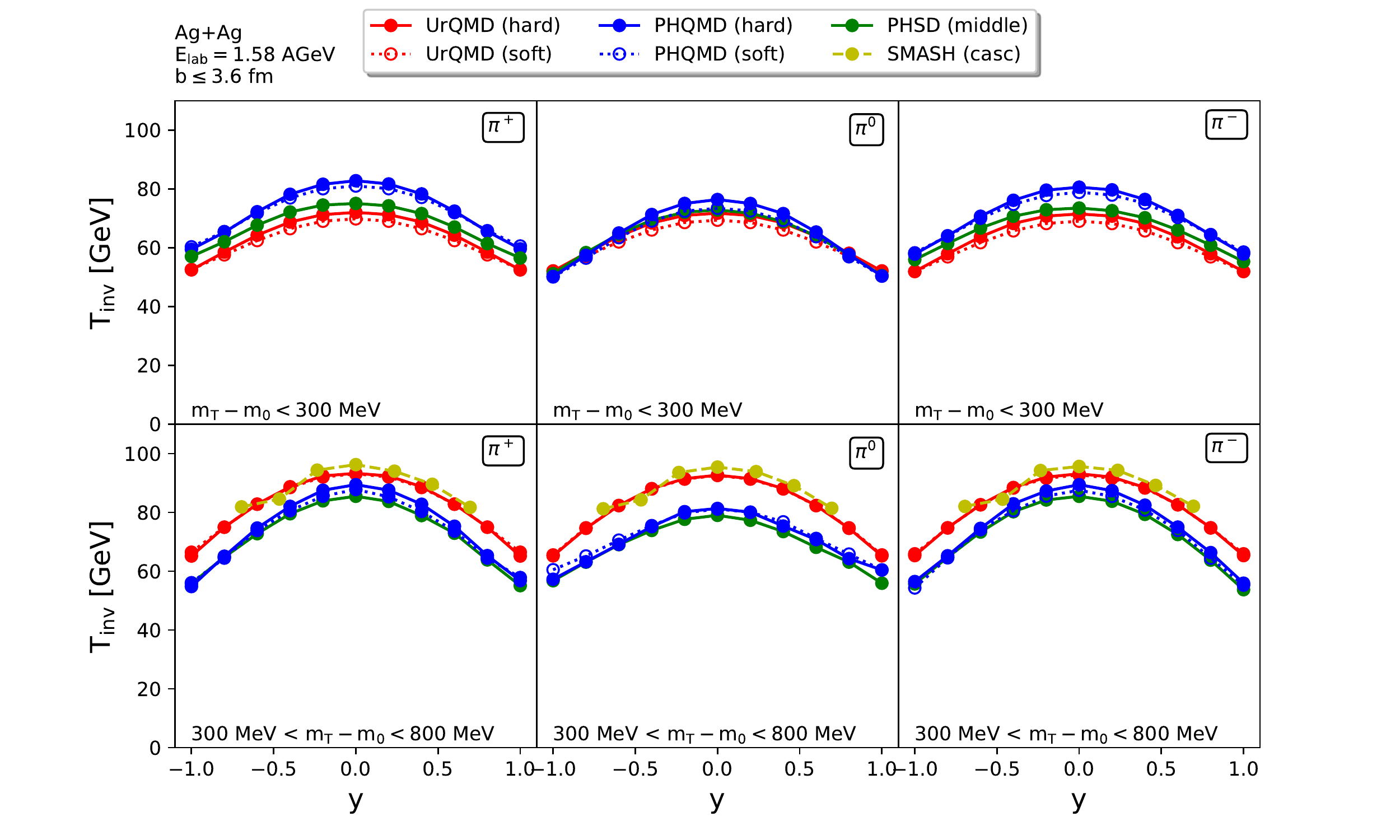}
    \caption{[Color online] Inverse slope parameter $T_\mathrm{inv}$ in dependence of the rapidity extracted by fitting the transverse mass spectra of $\pi^+$ (left), $\pi^0$ (center) and $\pi^-$ (right) with transverse mass below 300~MeV (top row) and with transverse mass between 300 and 800~MeV (bottom row) from UrQMD (red), PHSD (green), PHQMD (blue), SMASH (only high $m_\mathrm{T}$) \cite{Staudenmaier:2020xqr} (yellow). Open symbols and dotted lines refer to the soft EoS/no-medium interaction, while fulls symbols and full lines refer to the hard EoS/in-medium interaction and dashed lines refer to cascade mode}
    \label{fig:T_pions}
\end{figure}

Fig. \ref{fig:T_pions} (top row) shows the low $m_{\rm T}$ inverse slope parameters for $\pi^+$, $\pi^0$ and $\pi^-$ (from left to right) as a function of rapidity. The bottom part of the figure shows the corresponding inverse slopes for the high $m_\mathrm{T}$ region. The different models are shown as lines with symbols (open symbols and dotted lines refer to the soft EoS/no-medium interaction, while full symbols and full lines refer to the hard EoS/in-medium interaction and dashed lines refer to cascade mode): UrQMD (red), PHSD (green), PHQMD (blue) and SMASH (yellow). One generally observes that the low $m_{\rm T}$ inverse slopes are colder than the high $m_{\rm T}$ inverse slopes. This is to be expected from the production process of low $m_{\rm T}$ pions from the Delta decay. One also observes that all model calculations show consistent results ($T(y_{cm})=60\pm 10$~MeV) - the SMASH results can only be included in the high $m_{\rm T}$ comparison, due to the larger fit ranges employed in \cite{Staudenmaier:2020xqr}. At high transverse momenta the models do also show very similar behaviors and are consistent with ($T(y_{cm})=80\pm 10$~MeV).

\begin{figure} [t!hb]
    \includegraphics[width=\columnwidth]{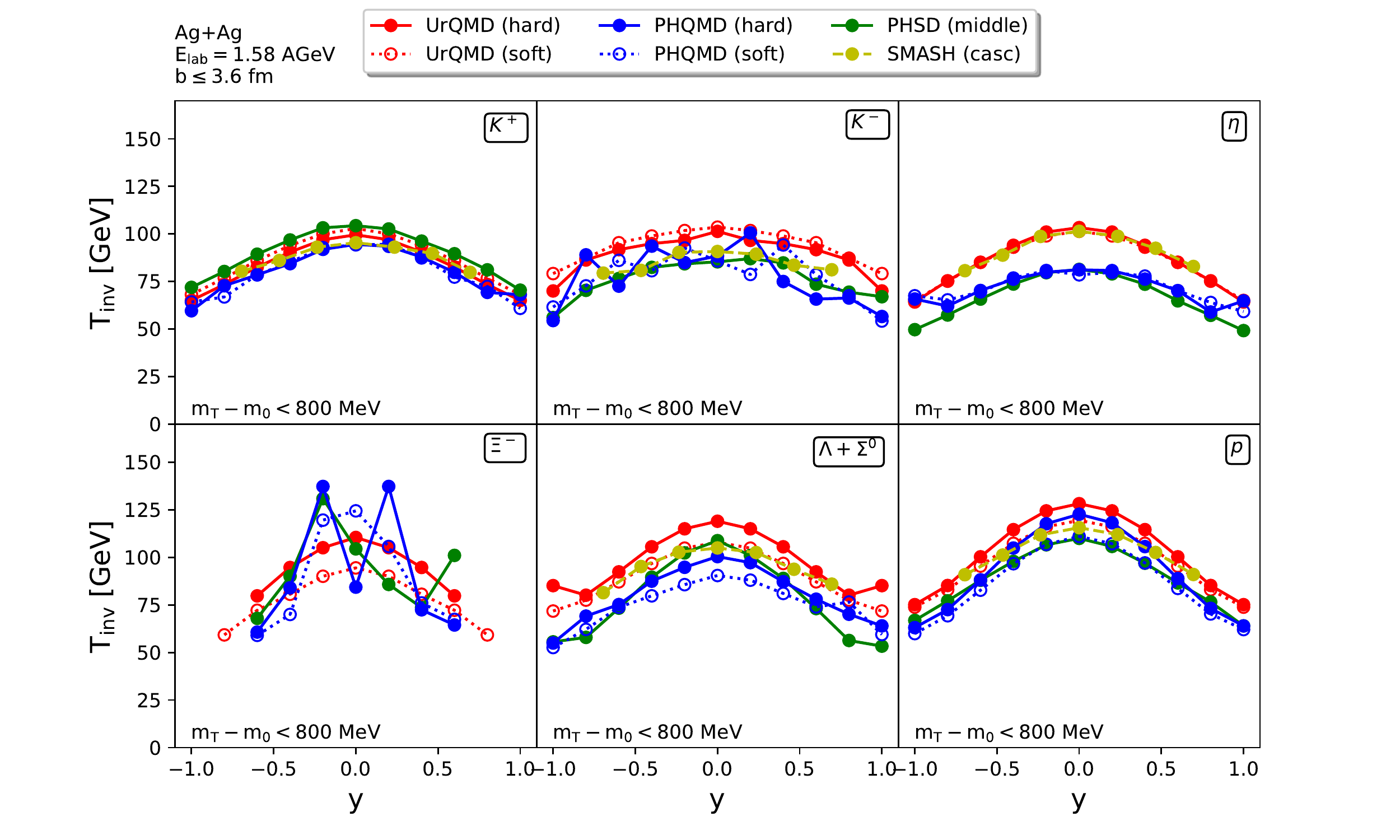}
    \caption{[Color online] Inverse slope parameter $T_\mathrm{inv}$ as a function of rapidity extracted by fitting the transverse mass spectra of protons (lower right), $K^+$ (upper left), $K^-$ (upper center), the $\eta$ (upper right), $\Lambda+\Sigma^0$ (lower center) and the $\Xi^-$ (lower left) in the transverse mass range $m_\mathrm{T}-m_0<0.8$~GeV from UrQMD (red), PHSD (green), PHQMD (blue), SMASH \cite{Staudenmaier:2020xqr} (yellow). Open symbols and dotted lines refer to the soft EoS/no-medium interaction, while fulls symbols and full lines refer to the hard EoS/in-medium interaction and dashed lines refer to cascade mode}
    \label{fig:T_hadrons}
\end{figure}

In Fig. \ref{fig:T_hadrons} we show the inverse slope parameters for protons (lower right), $K^+$ (middle left), $K^-$ (middle center) and $\Lambda+\Sigma^0$ (lower center) extracted from the transverse mass range $m_\mathrm{T}-m_0<0.8$~GeV as a function of rapidity. The symbols and line styles are the same as in the previous figure (UrQMD (red), PHSD (green), PHQMD (blue) and SMASH (yellow)).  

Finally, we address the question of the temperature of the emission source. Let us first explore the rapidity dependence of the temperature individually for each particle species. If the rapidity dependence of the inverse slopes can be described by a single fireball with a single temperature $T_\mathrm{eff}$  as given by
\begin{equation}\label{eq:T_eff}
    T_\mathrm{inv}(y) = \frac{T_\mathrm{eff}}{\cosh(y)},
\end{equation}
we can conclude that the system has reached thermal equilibrium to a high degree.

\begin{figure} [t!b]
    \includegraphics[width=\columnwidth]{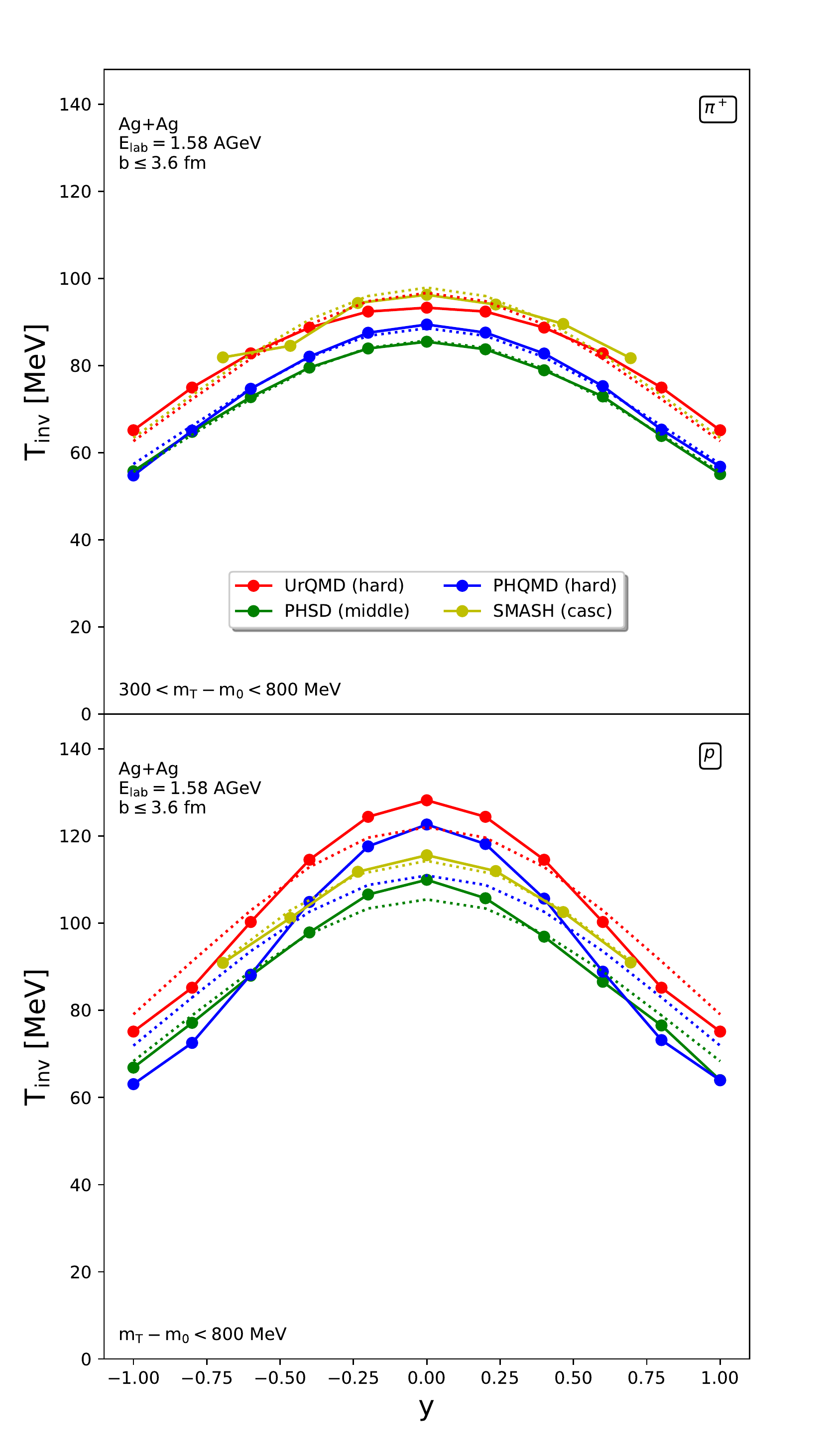}
    \caption{[Color online] Inverse slope parameter $T_\mathrm{inv}$ (solid line with circles) and the fit function Eq. \eqref{eq:T_eff} (dotted line) of $\pi^+$ in the transverse mass range $0.3<m_\mathrm{T}-m_0<0.8$~GeV (upper panel) and $p$ in the transverse mass range $m_\mathrm{T}-m_0<0.8$~GeV (lower panel) from UrQMD with hard EoS (red), PHSD with middle EoS (green), PHQMD with hard EoS (blue) and SMASH in cascade mode \cite{Staudenmaier:2020xqr} (yellow).}
    \label{fig:T_fits_thermal}
\end{figure}

Fig. \ref{fig:T_fits_thermal} shows exemplary fits for the high transverse mass pions and the protons for all investigated models. The inverse slope parameter $T_\mathrm{inv}$ (solid line with circles) and the fit function Eq. \eqref{eq:T_eff} (dotted line) of $\pi^+$ in the transverse mass range $0.3<m_\mathrm{T}-m_0<0.8$~GeV (upper panel) and protons in the transverse mass range $m_\mathrm{T}-m_0<0.8$~GeV (lower panel) from UrQMD with hard EoS (red), PHSD with middle EoS (green), PHQMD with hard EoS (blue) and SMASH in cascade mode (yellow) are shown. One sees that the assumption of a single fireball with a common temperature $T_\mathrm{eff}$ that is red shifted with rapidity provides a very good description of the simulation results. This might suggest that the source created in Ag+Ag reactions at such low energies has indeed achieved a large degree of thermalization and expands collectively.

The results for the effective temperatures $T_\mathrm{eff}$ of the source are summarized in Table \ref{tab:T_eff}.

\begin{table*}[h!tb]
    \centering
    \begin{tabular}{l||c|c|c|c|c|c|c|c|c|c|c|c}
        $T_\mathrm{eff}$ [MeV] & $\pi^+_{\mathrm{low}\,m_\mathrm{T}}$ & $\pi^0_{\mathrm{low}\,m_\mathrm{T}}$ & $\pi^-_{\mathrm{low}\,m_\mathrm{T}}$ & $\pi^+_{\mathrm{high}\,m_\mathrm{T}}$ & $\pi^0_{\mathrm{high}\,m_\mathrm{T}}$ & $\pi^-_{\mathrm{high}\,m_\mathrm{T}}$ & $K^+$ & $K^-$ & $\eta$ & $\Xi^-$ & $\Lambda+\Sigma^0$ & $p$ \\
        \hline
        \hline
        UrQMD (hard EoS)  & 75 & 75 & 75 & 97 & 96 & 97 &  99 & 105 & 102 & 103 & 116 & 122 \\
        UrQMD (soft Eos)  & 73 & 73 & 73 & 97 & 96 & 97 & 103 & 110 & 101 &  88 & 106 & 116 \\
        SMASH (casc)  & -  & -  & -  & 98 & 98 & 98 & 97  & 94  & 101 &   - & 105 & 114 \\
        PHSD (middle EoS, in-medium) & 79 & 75 & 78 & 86 & 81 & 86 & 106 &  90 &  79 & 101 &  94 & 106 \\
        PHQMD (hard EoS)  & 85 & 77 & 83 & 89 & 83 & 89 &  95 &  91 &  84 & 100 &  95 & 111 \\
        PHQMD (soft EoS)  & 85 & 75 & 82 & 87 & 84 & 87 &  94 &  92 &  84 &  97 &  87 & 103 \\
    \end{tabular}
    \caption{Effective temperature $T_\mathrm{eff}$ of the different hadron species extracted by fitting the inverse slope parameter with equation \eqref{eq:T_eff}. SMASH data was taken from \cite{Staudenmaier:2020xqr}}.
    \label{tab:T_eff}
\end{table*}

Obviously the effective temperatures for different particle species are different. This difference can be attributed to the transverse flow, which apparently leads to a mass dependent effective temperature. The flow effect can be best extracted using the high momentum tail of the distribution. (The effective source temperatures at low $m_{\rm T}$ are very similar in all models as they reflect essentially the decay kinematics of the Delta resonance.) The idea is that the true temperature of the source $T_\mathrm{source}$ is modified due to the transverse flow and can be extracted from $T_\mathrm{eff}$ in the limit $m_\mathrm{hadron}\rightarrow 0$. 

\begin{figure} [t!b]
    \includegraphics[width=\columnwidth]{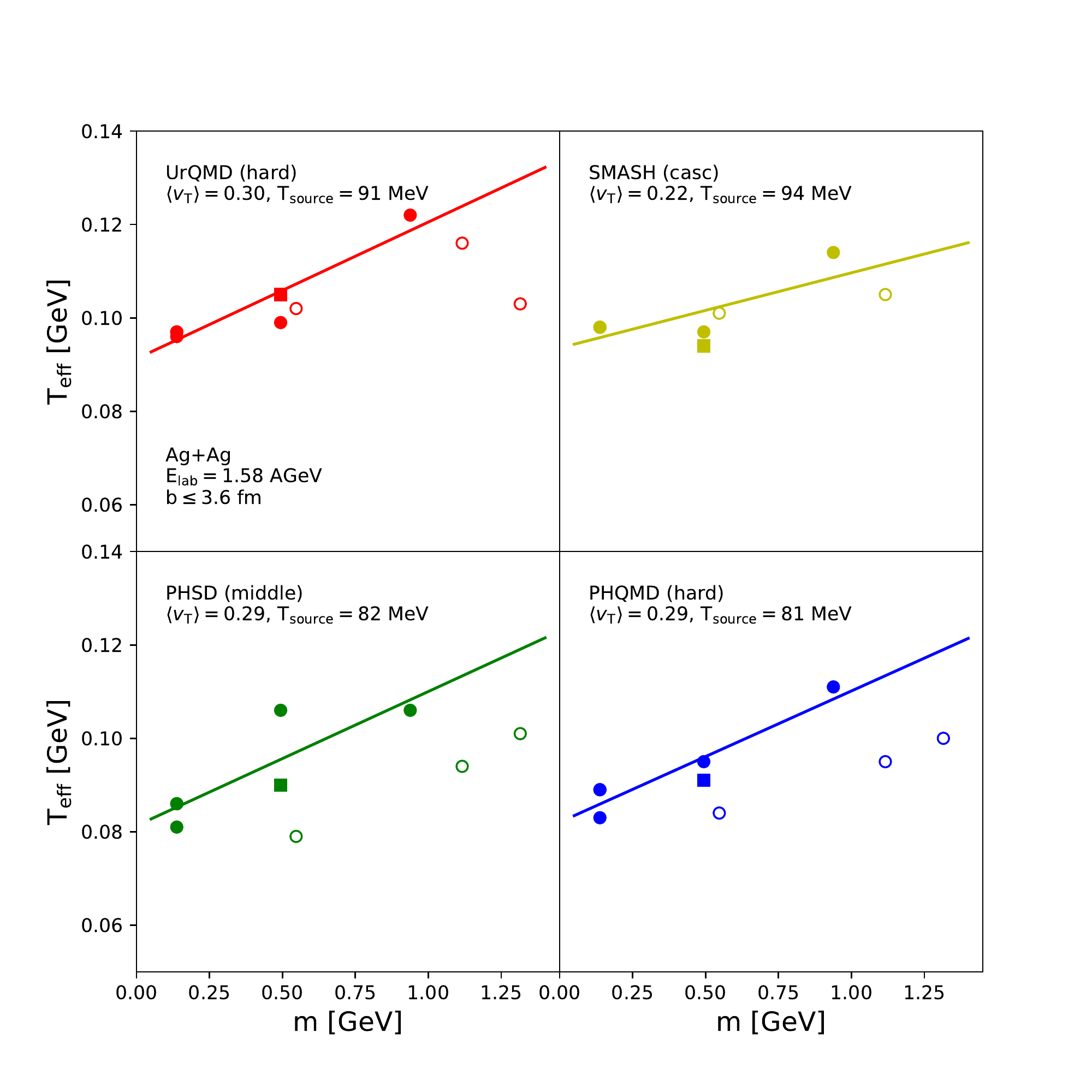}
    \caption{[Color online] Effective temperature $T_\mathrm{eff}$ of high $m_\mathrm{T}$ pions, $K^+$, $K^-$ (square), $\eta$, proton, $\Lambda+\Sigma^0$ and $\Xi^-$ as a function of the hadron mass from UrQMD with hard EoS (red, upper left), PHSD with middle EoS and in-medium interaction (green, lower left), PHQMD with hard EoS (blue, lower right) and SMASH in cascade mode \cite{Staudenmaier:2020xqr} (yellow, upper right). The line depicts a linear fit of the form $T_\mathrm{eff}=T_\mathrm{source}+\frac{1}{3} m \langle v_T\rangle^2$. The extracted average transverse expansion velocity, $\langle v_T\rangle=\sqrt{\langle v_T\rangle^2}$ and the source temperature $T_\mathrm{source}$ are depicted in the upper left corners of each plot.}
    \label{fig:Teff_m0}
\end{figure}

In Fig. \ref{fig:Teff_m0} we show the effective temperature $T_\mathrm{eff}$ of high $m_{\rm T}$ pions, $K^\pm$, $\eta$, proton, $\Lambda+\Sigma^0$ and $\Xi^-$ as a function of the hadron mass from UrQMD with hard EoS (red, upper left), PHSD with middle EoS and in-medium interaction (green, lower left), PHQMD with hard EoS (blue, lower right) and SMASH in cascade mode (yellow, upper right). The effective temperatures of all the hadrons are depicted as circles, despite the $K^-$ which is shown as a square to distinguish it from the $K^+$. The line depicts a linear fit of the form 
$$T_\mathrm{eff}=  T_\mathrm{source}+\frac{2}{3} \langle E_\mathrm{kin}\rangle=T_\mathrm{source}+\frac{1}{3} m \langle v_T\rangle^2.$$ 
The slope parameter $\langle v_T\rangle^2$ is fitted to the pion, (anti-)kaon and proton data (full symbols) while open symbols are not included in the fit. The extracted average transverse expansion velocity, $\langle v_T\rangle=\sqrt{\langle v_T\rangle^2}$ and the source temperatures $T_\mathrm{source}$ are depicted in the upper left corners of each plot.

We restrict the discussion to the hard EoS for brevity. First one observes that all models yield a similar source temperature $T_\mathrm{source}=80-95$~MeV when extrapolated to the zero flow limit, i.e. to $m=0$. This is in line with previous studies at HADES energies that have suggested source temperatures on the order of 80-100 MeV \cite{FOPI:1997xsg,Reichert:2020uxs} for similar system sizes and collision energies.

For non-zero masses UrQMD, PHSD, PHQMD and SMASH show substantial radial flow with average transverse flow velocities between $\langle v_T\rangle=0.22c - 0.3c$. The magnitude of the transverse flow velocity of UrQMD, PHSD, PHQMD and SMASH in Ag+Ag is smaller than the values obtained for Au+Au reactions (there a value of $\langle v_T\rangle=0.4c$ was extracted \cite{Harabasz:2020sei}), which is expected due to the smaller system size. Another interesting feature observed consistently in UrQMD, PHSD and in PHQMD is the smaller (as compared to the flow fit) effective temperature of Lambda's and $\Xi$'s. This feature was previously found in Ref. \cite{vanHecke:1998yu} and attributed to the early decoupling of of (multi-)strange baryons from the fireball and was also confirmed to exist in Refs. \cite{WA97:1998bwp,WA97:1998cdp}. 

\section{Summary}
In this study we compared a full breadth of established microscopic transport models to make predictions for the upcoming Ag+Ag data at $E_\mathrm{lab}=1.58$~$A$GeV ($\sqrt{s_\mathrm{NN}}=2.55$~GeV) by the HADES collaboration. We studied multiplicities, spectra and effective source temperatures of protons, $\pi^{\pm,0}$, $K^\pm$, the $\eta$, $\Lambda+\Sigma^0$ and the $\Xi^-$ using the UrQMD, PHSD, PHQMD and SMASH models which all rely on similar physics ideas, however, with substantial differences in their detailed implementation -- the different transport equations-of-motion - QMD versus BUU, potentials, realization of production mechanisms
and interactions, inclusion of in-medium effects etc.

Despite varying treatment of the dynamics encountered in heavy ion collisions the employed transport models all show a very nice agreement on the multiplicities of the investigated hadrons. The main differences are in the $\Xi^-$ production, which is treated drastically different between UrQMD/SMASH on one side employing high mass resonance states with explicit decays to $\mathrm{Resonance}\rightarrow \Xi+K+K$ in contrast to PHSD/PHQM which 
incorporate only non-resonant production channels for $\Xi$. A comparison of the spectra, summarized in effective source temperatures, reveals that all models show similar source temperatures around $T_\mathrm{source}=80-95$~MeV and average transverse flow around $\langle v_T\rangle=0.22c-0.3c$.

\begin{acknowledgements}
The authors thank Simon Spies, Jan-Hendrik Otto and Joachim Stroth from the HADES collaboration for fruitful discussion at the "Transport meets HADES, symposium during the HADES Collaboration Meeting XL" symposium and Szymon Harabasz from the HADES collaboration at the "CPOD2021 - International Conference on Critical Point and Onset of Deconfinement" conference. This article is part of a project that has received funding from the European Union’s Horizon 2020 research and innovation programme under grant agreement STRONG – 2020 - No 824093. Furthermore, we acknowledge support by the Deutsche Forschungsgemeinschaft 
(DFG, German Research Foundation): grant BR~4000/7-1 and   by the GSI-IN2P3 agreement under contract number 13-70. JS acknowledges support from the BMBF through the ErUM-data-pilot project as well as the Samson AG.
Also we thank the COST Action THOR, CA15213. 
Computational resources were provided by the Center for Scientific Computing (CSC) of the Goethe University
and the "Green Cube" at GSI, Darmstadt.

\end{acknowledgements}



\appendix

\section{Full transverse mass distributions}\label{app:trans_mass}
To provide the full informaton of the underlying transverse mass distribution we present the distributions for all models, all investigated particles species and for various rapidity bins in this appendix.

\begin{figure*} [t!hb]
    \centering
    \includegraphics[width=1.45\columnwidth]{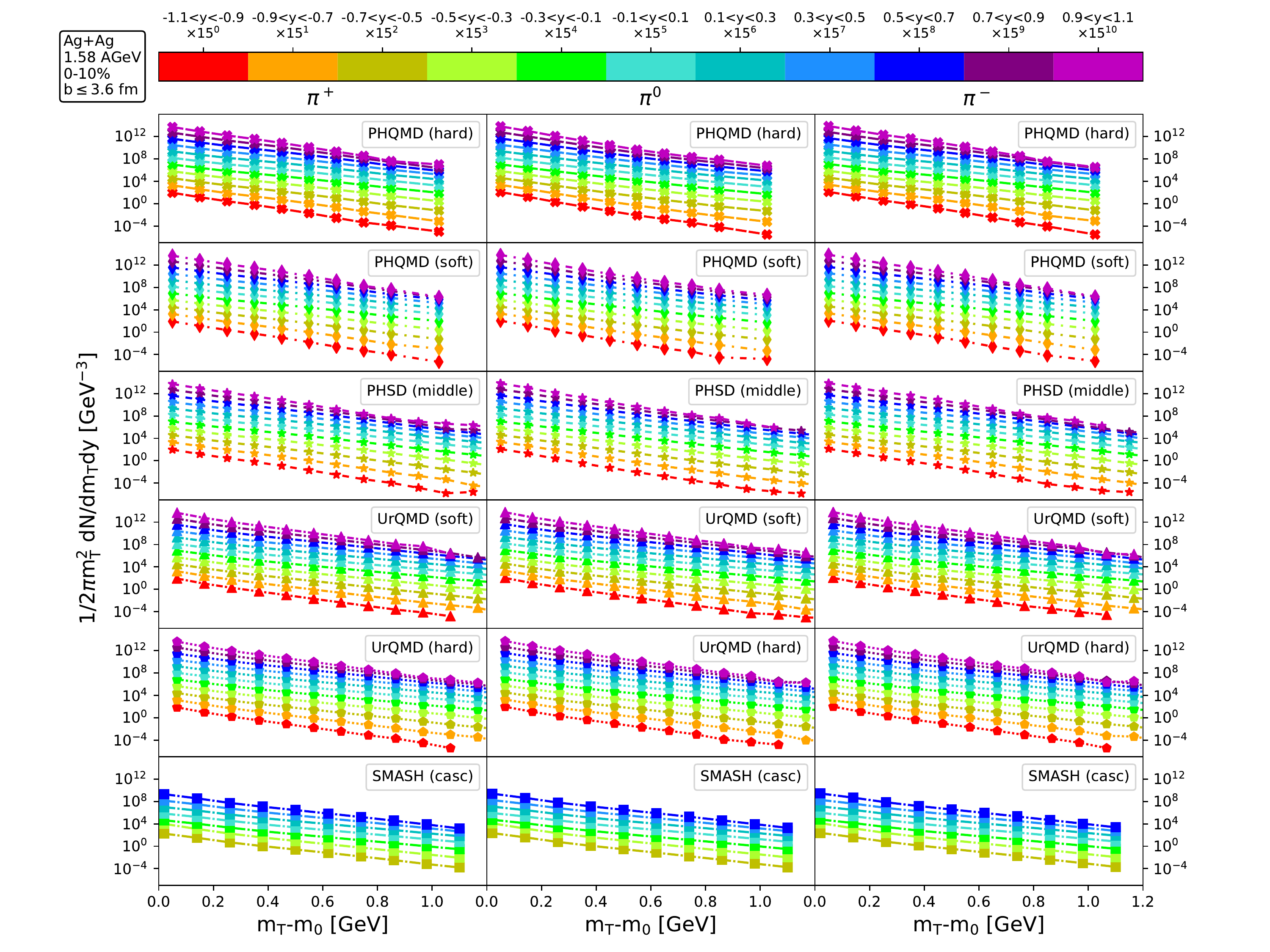}
    \caption{[Color online] Transverse mass spectra of $\pi^+$ (left), $\pi^0$ (middle) and $\pi^-$ (right) in different rapidity bins (see legend) in the 10\% most central Ag+Ag collisions at $E_\mathrm{lab}=1.58A$~GeV from UrQMD v3.5 with hard and soft EoS, PHSD with middle EoS and in-medium interaction, PHQMD with hard and soft EoS and SMASH in cascade mode. Note that the rapidity bins from the SMASH results \cite{Staudenmaier:2020xqr} are slightly different.}
    \label{fig:mT_spectra_pions}
\end{figure*}

In Fig. \ref{fig:mT_spectra_pions} we show the transverse mass spectra of $\pi^+$ (left), $\pi^0$ (middle) and $\pi^-$ (right) in different rapidity bins (see legend) in the 10\% most central Ag+Ag collisions at $E_\mathrm{lab}=1.58A$~GeV from UrQMD v3.5 with hard and soft EoS, PHSD with middle EoS and in-medium interaction, PHQMD with hard and soft EoS and SMASH in cascade mode. Note that the rapidity bins from the SMASH results \cite{Staudenmaier:2020xqr} are slightly different.

\begin{figure*} [t!hb]
    \centering
    \includegraphics[width=1.45\columnwidth]{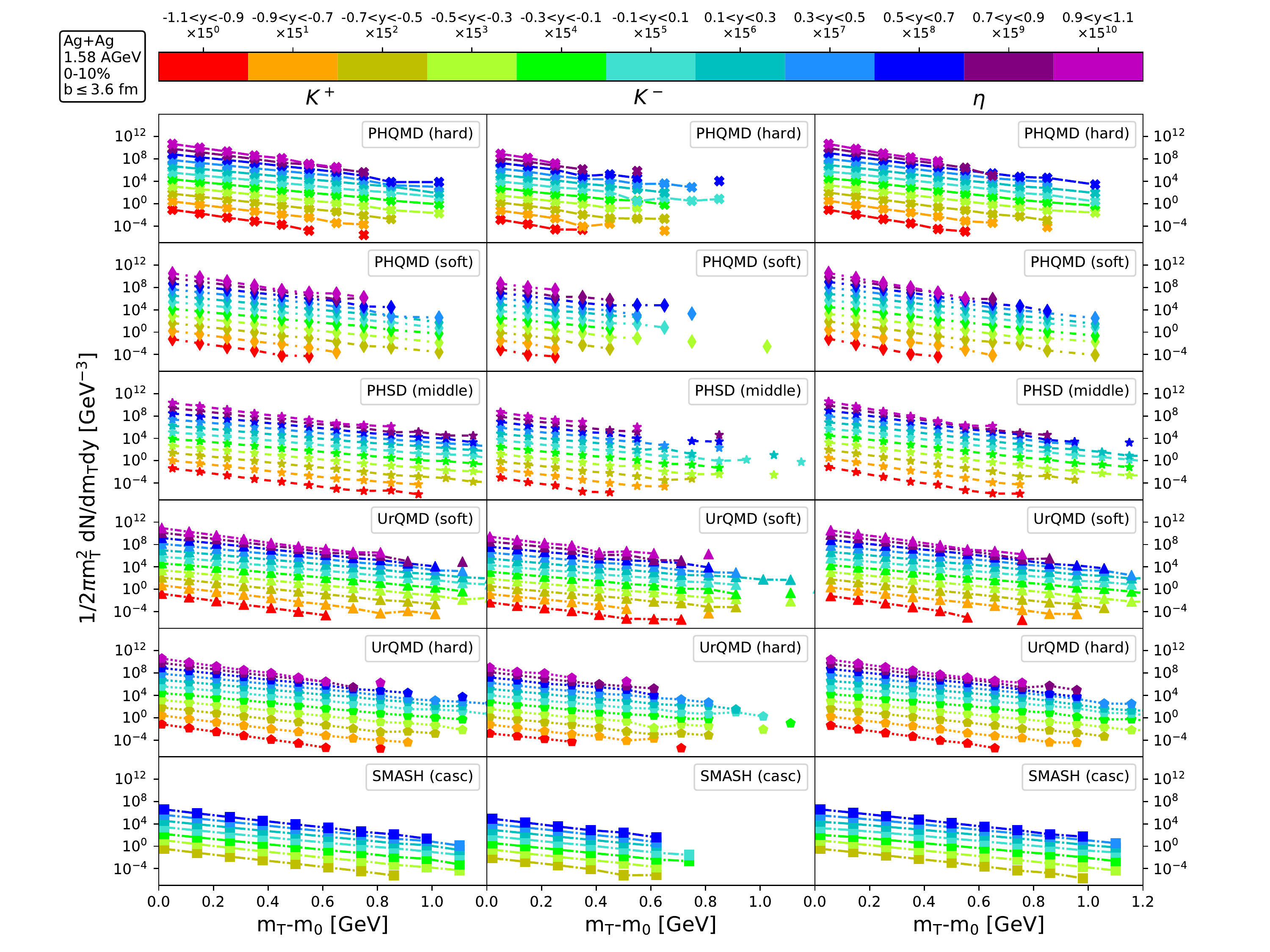}
    \caption{[Color online] Transverse mass spectra of $K^+$ (left), $K^-$ (middle) and $\eta$ (right) in different rapidity bins (see legend) in the 10\% most central Ag+Ag collisions at $E_\mathrm{lab}=1.58A$~GeV from UrQMD v3.5 with hard and soft EoS, PHSD with middle EoS and in-medium interaction, PHQMD with hard and soft EoS and SMASH in cascade mode. Note that the rapidity bins from the SMASH results \cite{Staudenmaier:2020xqr} are slightly different.}
    \label{fig:mT_spectra_mesons}
\end{figure*}

In Fig. \ref{fig:mT_spectra_mesons} we show the transverse mass spectra of $K^+$ (left), $K^-$ (middle) and $\eta$ (right) in different rapidity bins (see legend) in the 10\% most central Ag+Ag collisions at $E_\mathrm{lab}=1.58A$~GeV from UrQMD v3.5 with hard and soft EoS, PHSD with middle EoS and in-medium interaction, PHQMD with hard and soft EoS and SMASH in cascade mode. Note that the rapidity bins from the SMASH results \cite{Staudenmaier:2020xqr} are slightly different.

\begin{figure*} [t!hb]
    \centering
    \includegraphics[width=1.45\columnwidth]{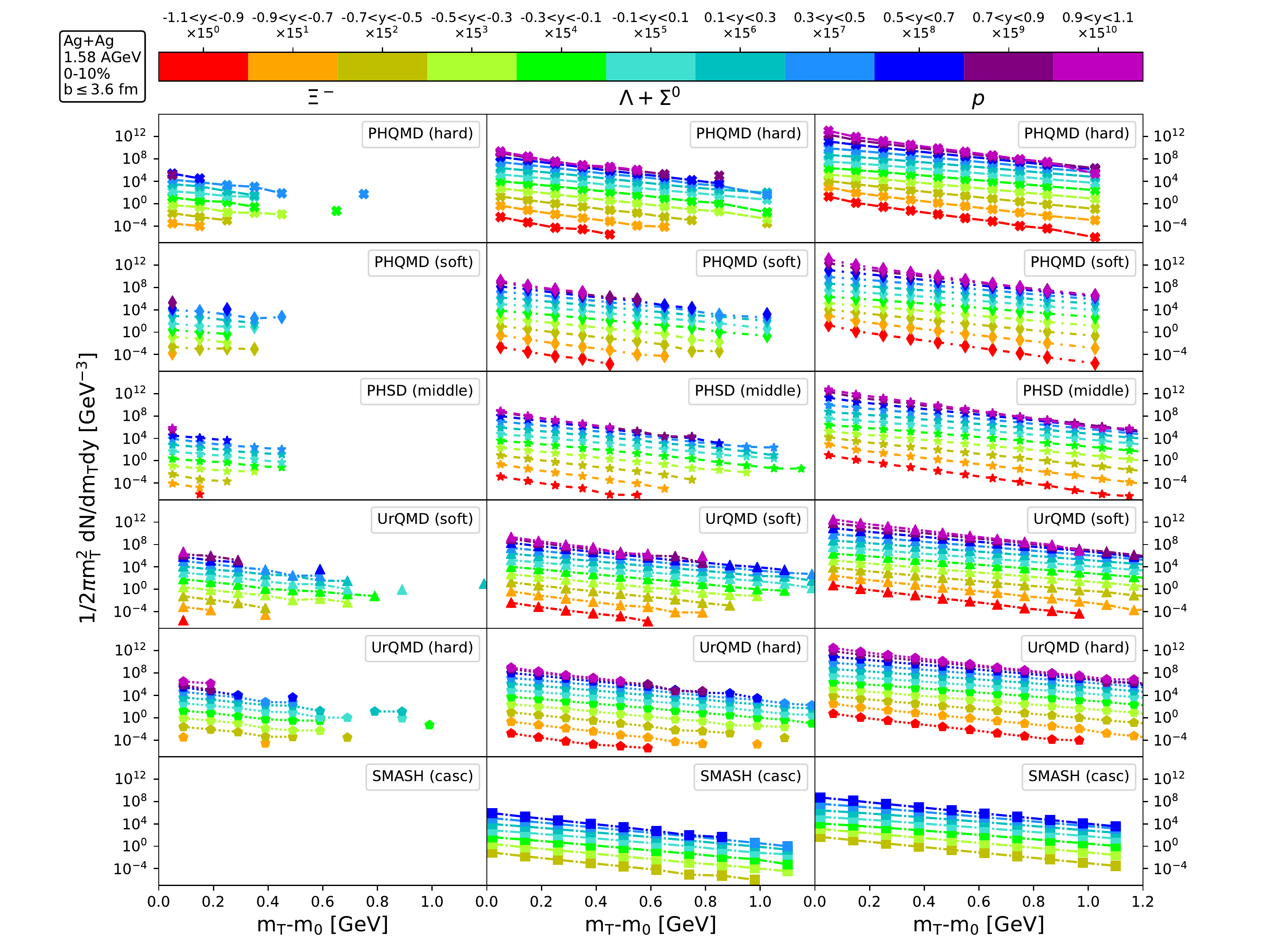}
    \caption{[Color online] Transverse mass spectra of $\Xi^-$ (left), $\Lambda+\Sigma^0$ (middle) and $p$ (right) in different rapidity bins (see legend) in the 10\% most central Ag+Ag collisions at $E_\mathrm{lab}=1.58A$~GeV from UrQMD v3.5 with hard and soft EoS, PHSD with middle EoS and in-medium interaction, PHQMD with hard and soft EoS and SMASH in cascade mode. Note that the rapidity bins from the SMASH results \cite{Staudenmaier:2020xqr} are slightly different.}
    \label{fig:mT_spectra_baryons}
\end{figure*}

And in Fig. \ref{fig:mT_spectra_baryons} we depict the transverse mass spectra of $\Xi^-$ (left), $\Lambda+\Sigma^0$ (middle) and $p$ (right) in different rapidity bins (see legend) in the 10\% most central Ag+Ag collisions at $E_\mathrm{lab}=1.58A$~GeV from UrQMD v3.5 with hard and soft EoS, PHSD with middle EoS and in-medium interaction, PHQMD with hard and soft EoS and SMASH in cascade mode. Note that the rapidity bins from the SMASH results \cite{Staudenmaier:2020xqr} are slightly different.

\end{document}